\documentclass[aps, reprint, nofootinbib]{revtex4-1}

\usepackage[sort&compress]{natbib}
\usepackage{amsmath}
\usepackage{graphicx}
\usepackage[rightcaption]{sidecap}
\usepackage{courier}
\usepackage{hyperref}
\usepackage{amsfonts}
\usepackage{color}
\usepackage{braket}
\usepackage{ulem}
\usepackage{subfig}
\usepackage{amsthm}
 
\theoremstyle{definition}

\newcommand{\be}{\begin{equation}}
\newcommand{\ee}{\end{equation}}

\newcommand{\lads}{\ell_\text{AdS}}

\pdfoutput=1
\begin{document}

\title{Large Breakdowns of Entanglement Wedge Reconstruction}
\author{Chris Akers}
\email{cakers@berkeley.edu}
\affiliation{Department of Physics, University of California, Berkeley, CA 94720, USA}
\author{Stefan Leichenauer}
\email{sleichen@gmail.com}
\affiliation{Alphabet (Google) X}
\author{Adam Levine}
\email{arlevine@berkeley.edu}
\affiliation{Department of Physics, University of California, Berkeley, CA 94720, USA}

\begin{abstract}
We show that the bulk region reconstructable from a given boundary subregion --- which we term the reconstruction wedge --- can be much smaller than the entanglement wedge even when backreaction is small. We find arbitrarily large separations between the reconstruction and entanglement wedges in near-vacuum states for regions close to an entanglement phase transition, and for more general regions in states with large energy (but very low energy density).
Our examples also illustrate situations for which the quantum extremal surface is macroscopically different from the Ryu-Takayanagi surface.
\end{abstract}

\maketitle

\section{Introduction}
Sometimes the gap between approximate and exact quantum error correction is more like a gulf. The application of quantum error correction to AdS/CFT first focused on the exact form of quantum error correction \cite{Almheiri:2014lwa, Dong:2016eik, Harlow:2016vwg}. An important result was that for every given CFT subregion $A$ together with some code subspace $\mathcal{H}_{\text{code}}$, there exists a maximal bulk subregion $a$ whose algebra of operators $\mathcal{A}_a$ acting on $\mathcal{H}_{\text{code}}$ can be encoded in the algebra of boundary operators $\mathcal{A}_A$. 

Furthermore, it was discovered that the region $a$ is precisely the bulk region bounded by $A$ and its Ryu-Takayanagi (RT) surface at leading order in the $G_N$ expansion \cite{Ryu:2006ef, Ryu:2006bv}. The Ryu-Takayanagi surface is defined as the minimal area extremal surface homologous to the boundary region $A$ whose area in Planck units gives the CFT entropy of the boundary region $A$ to leading order in the $G_N$ expansion. The bulk region $a$ is referred to as the entanglement wedge. 

An essential input for the reconstructability of the entanglement wedge bounded by the RT surface was the equality between bulk and boundary relative entropies first discussed in \cite{Jafferis:2015del}. Namely, 
\begin{align}
S(\rho_A || \sigma_A) = S(\rho_a || \sigma_a) + \mathcal{O}(G_N)
\end{align}
where $\rho_a$ is any bulk state on the entanglement wedge whose bulk geometry is perturbatively close in $G_N$ to that of $\sigma_a$. This equality, however, is only true up to next-to-leading order in $G_N$. 

Naively, one might think that accounting for these higher order corrections affects the results of entanglement wedge reconstruction only when one asks about questions sensitive to higher and higher orders in $G_N$. 

Importantly, however, at higher orders in $G_N$, the boundary entropy is no longer given by the area of the RT surface but instead by the generalized entropy associated to the quantum extremal surface \cite{Engelhardt:2014gca}. This means that the entanglement wedge, defined now as the bulk region bounded by the minimal generalized entropy quantum extremal surface, is defined in a state-dependent way. Hayden \& Penington \cite{Hayden:2018khn} first noticed that this state-dependence leads to an important modification of the leading order entanglement wedge reconstruction theorems.\footnote{There may be a way to interpret the argument of \cite{Dong:2016eik} that gives the same prescription for the reconstructable region that \cite{Hayden:2018khn} does. It is to restrict the algebra of operators to only those in the entanglement wedge for every state in the code subspace {\it including mixed states}. However, this prescription does not obviously follow from the argument in \cite{Dong:2016eik} (which implies that considering mixed states is unimportant) and obscures the importance of the non-zero difference between bulk and boundary relative entropies.}
They found that the bulk region that is reconstructable given access only to the boundary region $A$ is not the entanglement wedge, but is instead a potentially much smaller region that depends on the code subspace.
We call this region the reconstruction wedge, which we now define for code subspaces in which all geometries are perturbatively-close.\footnote{We can generalize this definition for code spaces that include states with non-perturbatively different geometries as follows. For each state, we can define the algebra of operators in the entanglement wedge in the bulk using some non-perturbative gauge-invariant prescription. Taking the intersection of those algebras over all states in the code subspace defines the reconstruction wedge algebra.}

\vspace{3mm}
\noindent{\bf Definition:}
The {\it reconstruction wedge} of boundary region $A$ is the intersection of all entanglement wedges of $A$ for every state in the code subspace, pure or mixed.

\vspace{3mm}

One might think that the reconstruction wedge differs from the entanglement wedge of a given pure state in the code subspace only at subleading order as long as all states in the subspace have similar geometries.
After all, the quantum extremal surfaces for states in the same code subspace are naively different from the RT surface only at sub-leading orders in $G_N$.
Hayden \& Penington \cite{Hayden:2018khn} demonstrated that for large enough code subspaces this intuition is false. 

Consider trying to reconstruct a black hole operator using a boundary region $A$ that is barely more than half the CFT.
If this black hole were in a pure state, the entanglement wedge of $A$ would include the black hole. 
However, if our code subspace included, say, all $e^{S_\text{BH}}$ black hole microstates, then the reconstruction wedge is much smaller. 
Indeed, one state in our code subspace is the maximally-mixed state, whose entanglement wedge excludes the black hole! 
Therefore, the reconstruction wedge excludes the black hole.

Of course, $A$ will be able to reconstruct black hole operators on a restricted code subspace.
If there are few enough states in the subspace, then there is no state for which the black hole entropy can shift the dominant RT surface.
This type of reconstruction is ``state-dependent'' (really subspace-dependent) in the sense that, while there might be no one reconstruction procedure that works on a particular large code subspace, there are reconstruction procedures that work on smaller subspaces.

State-dependent reconstruction has been historically associated with black hole physics, particularly in the context of the firewall paradox \cite{Papadodimas:2012aq}. The primary goal of this work will be to demonstrate that the phenomena of Hayden \& Penington \cite{Hayden:2018khn}, as well as \cite{Penington:2019npb}, can be decoupled from the presence of horizons. In particular, we will find a family of simple code subspaces without black holes which exhibit macroscopic gaps between the pure state entanglement wedges and the reconstruction wedge. This signals the necessity of subspace-dependent bulk reconstruction even in the absence of black holes.

The states we consider consist of collections of particles in the bulk with internal degrees of freedom, i.e., spin states. All of the states in our code subspaces will be exactly the same in terms of the bulk energy distribution. This means they will all have exactly the same geometry, which simplifies the analysis. The states in the code subspace differ only by the internal states of the particles. The maximal bulk entropy is proportional to the number of particles. We will consider examples with only a few bulk particles, as well as examples with a very large number, of order $G_N^{-1}$, in which case they will be treated as a cloud of pressureless dust to account for backreaction. In all cases, we can explicitly calculate all of the geometric quantities of interest and we also have a complete understanding of the quantum states involved, unlike the case of a black hole.

We hope that this simplicity will allow us to abstract the subtleties of state-dependent reconstruction away from the black hole arena and into more familiar territory.
Our setups are also of independent interest as examples in which the position of the quantum extremal surface is macroscopically distinct from that of the classical RT surface. Such situations have recently become important in the study of reconstructing black hole interiors \cite{Penington:2019npb,Almheiri:2019psf}.
We begin in Section II by describing near-vacuum states --- containing only a few particles --- which exhibit large gaps between the entanglement and reconstruction wedges. In Section III we analyze the case of dustballs, which are states with a large number of bulk particles (and hence a large amount of energy), but with small energy density and well-controlled backreaction. In Section IV, we end with a discussion of our results.

\section{Unreconstructable regions of the entanglement wedge}
\begin{figure*}
   \begin{center} 
    \subfloat[]{\includegraphics[width=\columnwidth]{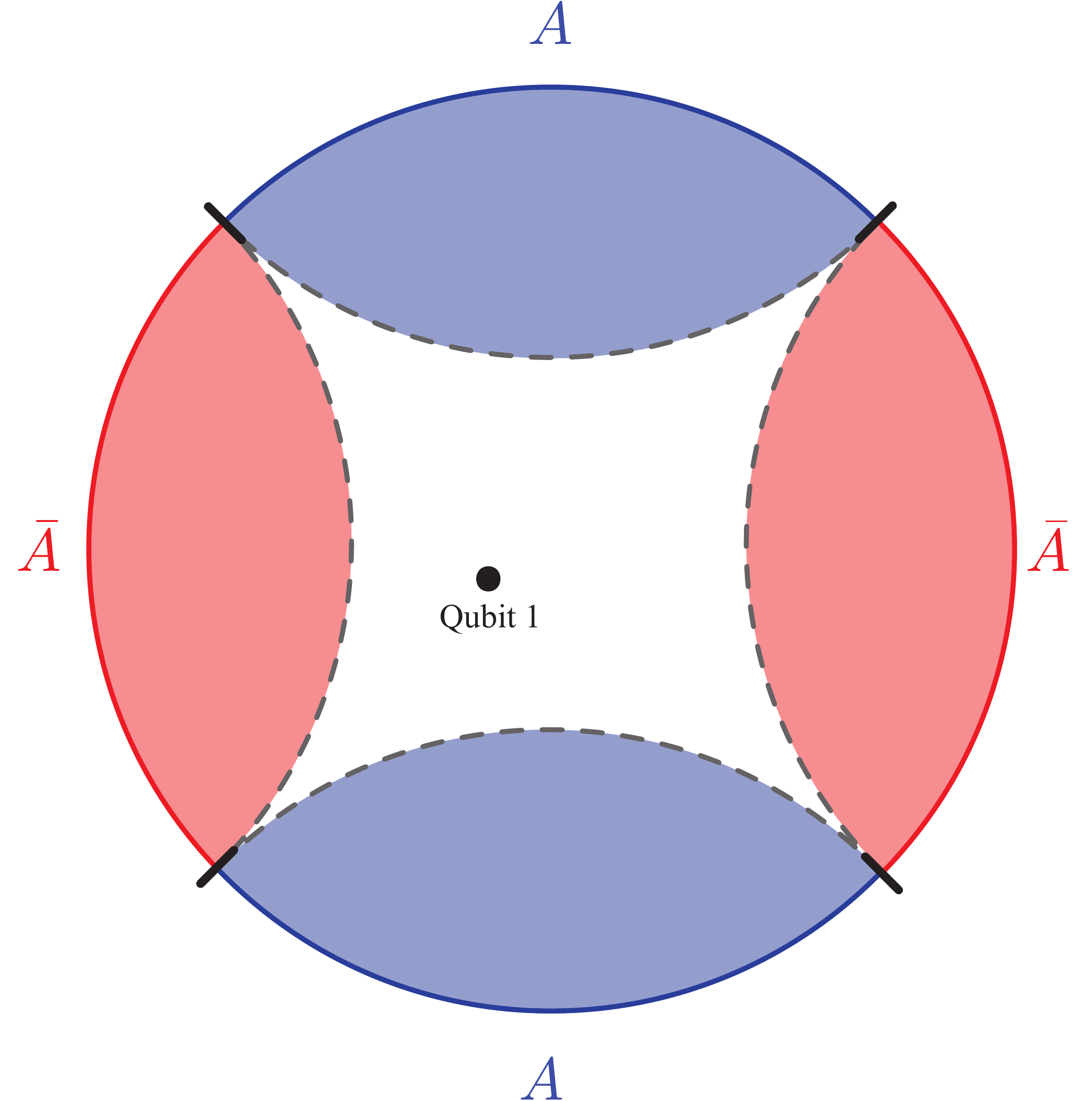}}
    \hfill
    \subfloat[]{\includegraphics[width=\columnwidth]{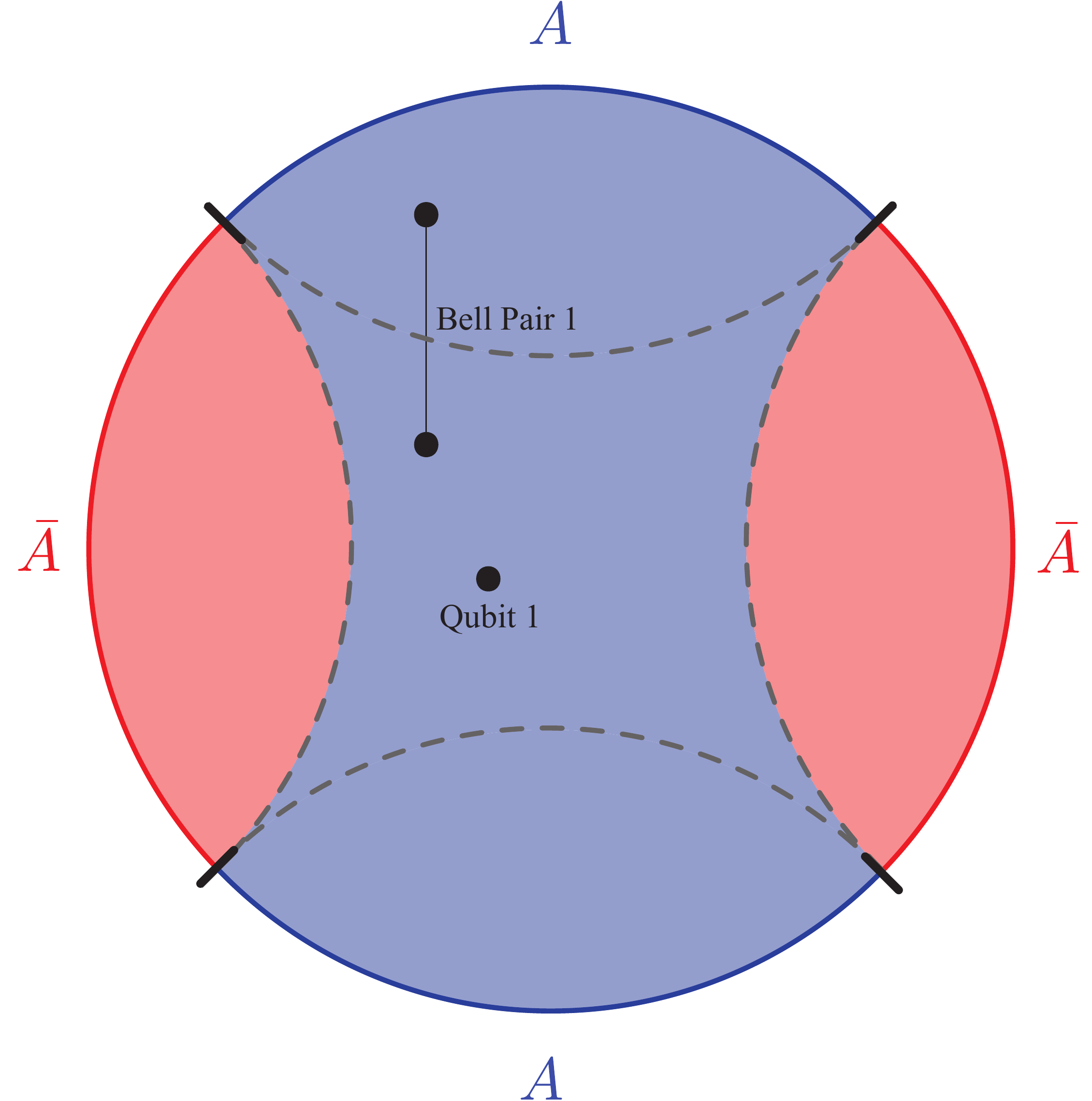}}
    \end{center}
    \caption{In blue is the reconstruction wedge of $A$ (blue boundary region). In red is the reconstruction wedge of $\bar{A}$ (red boundary region). Both $A$ and $\bar{A}$ are defined by dividing the boundary into four equal pieces and taking the union of two disconnected components. The quantum extremal surface of each connected component is depicted as thin dashed gray lines. (a) Neither reconstruction wedge includes the white region (the ``middle'') in the two-dimensional code subspace corresponding to the state Qubit 1 on top of the vacuum. (b) Adding a Bell pair between the entanglement wedge of one of $A$'s connected components and the middle causes the reconstruction wedge of $A$ to include the middle.}
    \label{fig:setup1and2}
\end{figure*}

We start with a simple example to illustrate the difference between the entanglement wedge (EW) and the reconstruction wedge (RW). 
Consider the vacuum state of $2{+}1$d AdS, and divide the boundary circle into four equally-sized intervals. 
There is a region of the bulk outside the EW of each of these four regions --- we call this the ``middle'' of the bulk.
See Figure \ref{fig:setup1and2}.
Let $A$ be the union of two disconnected intervals, and $\bar{A}$ the union of the other two. 
$A$ and $\bar{A}$ sit on an entanglement entropy phase transition: there are two candidate RT surfaces with equal areas and bulk entropies, and thus it is ambiguous whether the EW of $A$ should include the middle.

Now consider placing a single qubit in the middle of the bulk. Call this Qubit 1.
If Qubit 1 is in a mixed state, its entropy breaks the degeneracy betweeen the two candidate RT surfaces.\footnote{Technically, we should include the gravitational dressing of this qubit to the boundary. This dressing will have to cross at least one RT surface and change its area in Planck units by an amount comparable to the qubit's entropy. However, we can safely ignore this dressing by, for example, dressing the qubit equally to $A$ and $\bar{A}$, such that it changes the areas equally and does not affect the degeneracy.} 
In that case the EWs of both $A$ and $\bar{A}$ will exclude the middle.

To discuss the implications for the RW in this simple scenario, we should first define our code subspace. For now we consider a two-dimensional code subspace consisting of AdS with Qubit 1 in the the middle. All states in this code subspace have the same geometry, and differ only by the state of Qubit 1. To compute the RW of reigon $A$, we must compute the EW of $A$ for all states in the code subspace, {\it including mixed states}.
The RW is the intersection of all of these EWs.
In this case that intersection is equal to the EW corresponding to the most entropic state of Qubit 1.\footnote{We believe that the RW in general equals the EW of the most entropic bulk state. This would be interesting to prove, but our results do not depend on this.} 
Thus the middle is excluded from the RW of $A$, and similarly from $\bar{A}$.\footnote{Here, the difference between the EW and RW is AdS-scale. We could have made the difference arbitrarily large by dividing the boundary into $n$ equally sized intervals instead of four, letting $A$ be the union of every other connected component. Then the difference in volume between the RW and EW would scale linearly with $n$ for large $n$.}
See Figure \ref{fig:setup1and2}(a).

We can easily modify this example to a situation where $A$ and $\bar{A}$ are not on an entanglement phase transition such that we can see a macroscopic difference between the RW and EW. First consider the effect of adding a Bell pair to the bulk as follows. In addition to Qubit 1, place two additional qubits in the bulk which are maximally entangled with each other. One of these new qubits is placed in the middle of the bulk, and the other is placed well inside the EW of one of the connected components of $A$. See Figure \ref{fig:setup1and2}(b). These two new qubits form Bell Pair 1. The code subspace is still two-dimensional, corresponding to bulk states where the three qubits are at fixed positions and Bell Pair 1 is always in the same state, while the state of Qubit 1 remains arbitrary. Because of the presence of Bell Pair 1, the EW of $A$ includes the middle for all states in this code subspace (both pure and mixed). Hence the RW of $A$ also includes the middle, and coincides with the EW of $A$ for all states in the code subspace.

Now we add yet another qubit to the middle of the bulk, called Qubit 2, and we enlarge the code subspace to consist of the four-dimensional state space of Qubits 1 and 2. The EW of $A$ will include the middle of the bulk only for states in which Qubit 1 and 2 have no more than $\log 2$ von Neumann entropy, which is the entropy cost that must be paid to cut Bell Pair 1 when the middle is excluded. Therefore, the RW of $A$ will once again exclude the middle of the bulk.
This serves as our first example where the RW is macroscopically smaller than the EW, and also where complementary recovery exhibits a macroscopic breakdown.\footnote{There is an important caveat to all of these examples in which the difference in generalized entropy between two candidate quantum extremal surfaces is $\mathcal{O}(1)$. The fluctuations in area are naively of order $\delta A \sim \sqrt{\ell G_N}$ due to graviton fluctuations, where $\ell$ is a characteristic length scale of the state in question. This would lead to fluctuations $\frac{\delta A}{4G_N} \sim \sqrt{\ell/G_N}$, \cite{Engelhardt:2014gca} which would be much larger than the difference in entropy in the limit $G_N \to 0$. Hence it could be argued that it is unclear whether $A$ can really reconstruct the middle. This would be interesting to study in future work. We could mitigate this issue by studying states of fixed RT area like those studied in \cite{Dong:2018seb, Akers:2018fow}. We thank Geoff Penington for bringing this to our attention.}

\begin{figure}
    \includegraphics[width=\columnwidth]{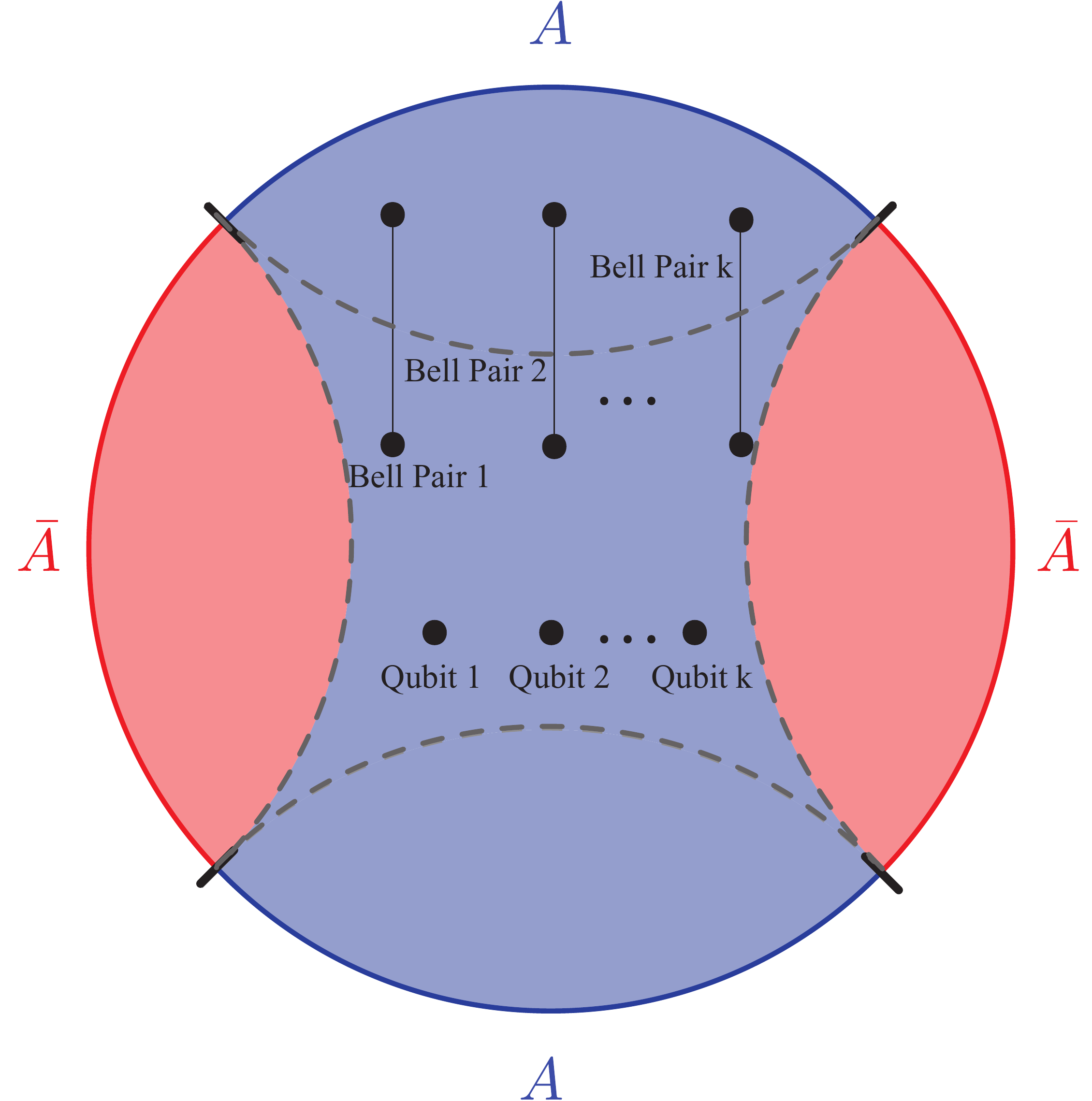}
    \caption{In blue is the reconstruction wedge of $A$ (blue boundary region). In red is the reconstruction wedge of $\bar{A}$ (red boundary region). $A$ can reconstruct Qubit $1$ through Qubit $k$ provided there are at least $k$ Bell pairs entangling the middle and the entanglement wedge of one of the connected components of $A$.}
    \label{fig:setup3}
\end{figure}

We can modify the example once again by adding a Bell Pair 2 in the same way as we added Bell Pair 1 to make the RW of $A$ include the middle. The general pattern is clear: if we construct a code subspace out of $k$ qubits in the middle, then $A$'s RW will only include the middle if we have at least $k$ Bell Pairs to compensate. See Figure \ref{fig:setup3}. In all of these cases the RW of $\bar{A}$ does not include the middle.

A similar set of examples can be constructed without the Bell Pairs by changing the relative sizes of $A$ and $\bar{A}$. Returning to the vacuum state, consider enlarging the two components of $A$ slightly, such that the minimal quantum extremal surface is the one nearer $\bar{A}$. Let the difference in area between the two competing surfaces be $\Delta \mathcal{A}$.
Now again place $k$ qubits in the middle (the exact location of the middle has also shifted slightly in this example), and let the code subspace be the $2^k$-dimensional state space of these qubits.
The RW of $A$ will only include the middle if $k \log 2 < \Delta \mathcal{A} / 4 G_N$.

As long as $\Delta \mathcal{A} \sim \mathcal{O}(G_N)$ there should be no difficulty placing $k \sim \Delta \mathcal{A}/G_N$ qubits in the middle of the bulk. However, if $\Delta \mathcal{A} \sim \mathcal{O}(G_N^0)$ then the backreaction of those $k$ qubits would cause a black hole to form. In the next section we will show how to construct a different set of examples where the backreaction in the bulk remains small, yet we can still explicitly define code subspaces of dimension $2^k$ where $k = \mathcal{O}(G_N^{-1})$ that exhibit macroscopic separation between the EW and RW.

\section{The Dustball}

\subsection{Geometry}

\begin{figure}
    \centering
    \includegraphics[width=.5\columnwidth]{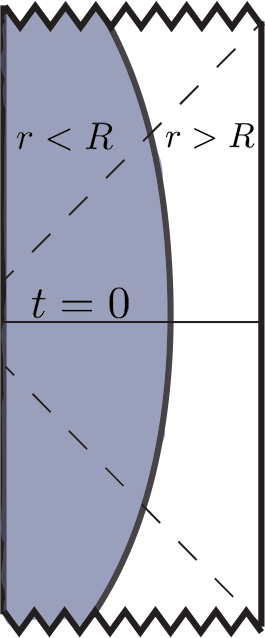}
    \caption{The Penrose diagram of the Oppenheimer-Snyder collapse. In the range $r < R$, the metric is hyperbolic FRW. For $r > R$, it is BTZ. The horizons do not reach the $t = 0$ surface.}
    \label{fig:OScollapse}
\end{figure}

Consider the Oppenheimer-Snyder collapse of a spherical ball of pressureless dust of uniform density in $2{+}1$d AdS. 
We refer to this as the dustball.
We take $t = 0$ to be the moment of time symmetry, where the dustball has maximum radius. 
Within the dustball, the spacetime is described by a portion of an FRW universe with past and future singularities. 
The outside is described by a BTZ black hole.
See Figure \ref{fig:OScollapse}.

We are only concerned with the state at $t = 0$.
At this time the energy density is $\rho_0$, and we will take $\rho_0$ to be very small in the sense that
\begin{align}
    \eta \equiv 8 \pi G_N \rho_0 \ell^2_{\text{AdS}} \ll 1~.
\end{align}
The spatial part of the metric outside the dustball at $t = 0$ is given by 
\begin{equation}
    ds^2_{\text{out}} = \frac{\lads^2}{r^2 - r_h^2} dr^2 + r^2 d\phi^2,~~ r > R,
\end{equation}
while inside the dustball it is
\begin{equation}
    ds^2_{\text{in}} = \frac{1}{1 + a_0^{-2} r^2} dr^2 + r^2 d\phi^2,~~ r < R.
\end{equation}
Here $a_0$ is the scale factor of the interior FRW universe at its maximum size. From the Friedmann equation we have
\begin{equation}
    \ell^2_{\text{AdS}} a_0^{-2} = 1 - \eta,
\end{equation}
and by demanding continuity of the metric at $r = R$ we learn
\begin{equation}
    \lads^2 + r_h^2 = \eta R^2~.
\end{equation}
We are interested in a large dustball, meaning
\begin{equation}
    R \gg r_h \gg \lads~.
\end{equation}
In particular, this means that outside the dustball the metric is near-vacuum:
\begin{equation}
    ds^2_{\text{out}} = \frac{\lads^2}{r^2 + \lads^2}\left(1 + \mathcal{O}(\eta^2)\right) dr^2 + r^2 d\phi^2,~~ r > R.
\end{equation}
This makes the analysis of extremal surfaces outside of the dustball especially simple, which we will take advantage of below.

\subsection{The Code Subspace}
Every state in the code subspace is going to have exactly the same dustball geometry described above. The only degrees of freedom are the internal states of the dust particles, which are completely degenerate in terms of energy. We can think of these degrees of freedom as representing the spins of the dust particles, and the total code subspace thus has dimension $s^k$, where $k$ is the number of dust particles and $s$ is the number of spin states for each particle.

We note that there should be no issue with assuming the internal states do not backreact at the order we care about.
The relevant energy scale is set by the total mass, which is proportional to the number of particles.
On the other hand, any spin-spin interactions are suppressed by $G_N$ (because to leading order the particles are free) as well as the distance between the particles. 
That distance is large because we are assuming the density of particles is very small. 
Hence we can safely work in a regime in which the energy difference between different spin states of the dustball is negligible relative to its mass. 
The exterior geometry corresponds to BTZ of the same mass for each of the different states. 

\subsection{Arbitrarily Large Unreconstructable Regions}
In the dustball code subspace we can construct examples with large separations between reconstruction and entanglement wedges.
Fix $G_N$ small but finite so that the bulk is well-described by semiclassical physics.
The largest entropy state of the dustball is that in which each particle's internal state is maximally mixed.
In that case the dustball has a constant entropy density, and the total entropy is linear in the dustball radius $R$ when the radius is large.
We are ultimately interested in choosing $R$ to be large enough that the total entropy is $\mathcal{O}(G_N^{-1})$.

We will now define regions $A$ and $\bar{A}$ of the boundary such that the area difference $\Delta \mathcal{A}$ between the areas of the quantum extremal surface nearer $A$ and that nearer $\bar{A}$ is $\mathcal{O}(G_N^0)$. 

Divide the boundary into $2n + 1$ equally-sized subregions, for $n$ an integer.
We note $n$ should be large enough such that the quantum extremal surface of the union of any two of these subregions remains entirely outside the dustball. This will require $n\sim R/\lads$. Label the intervals $1,2, \ldots, 2n+1$ in order around the boundary. Let $A$ be the union of all of the odd-numbered intervals. Note that interval $1$ and interval $2n+1$ are adjacent. The complement $\bar{A}$ consists of all of the even-numbered intervals, none of which are adjacent. Said another way, both $A$ and $\bar{A}$ consist of $n$ disjoint intervals, but one of the intervals in $A$ is twice as large as all of the others.

\begin{figure*}
    \subfloat[]{\includegraphics[width=\columnwidth]{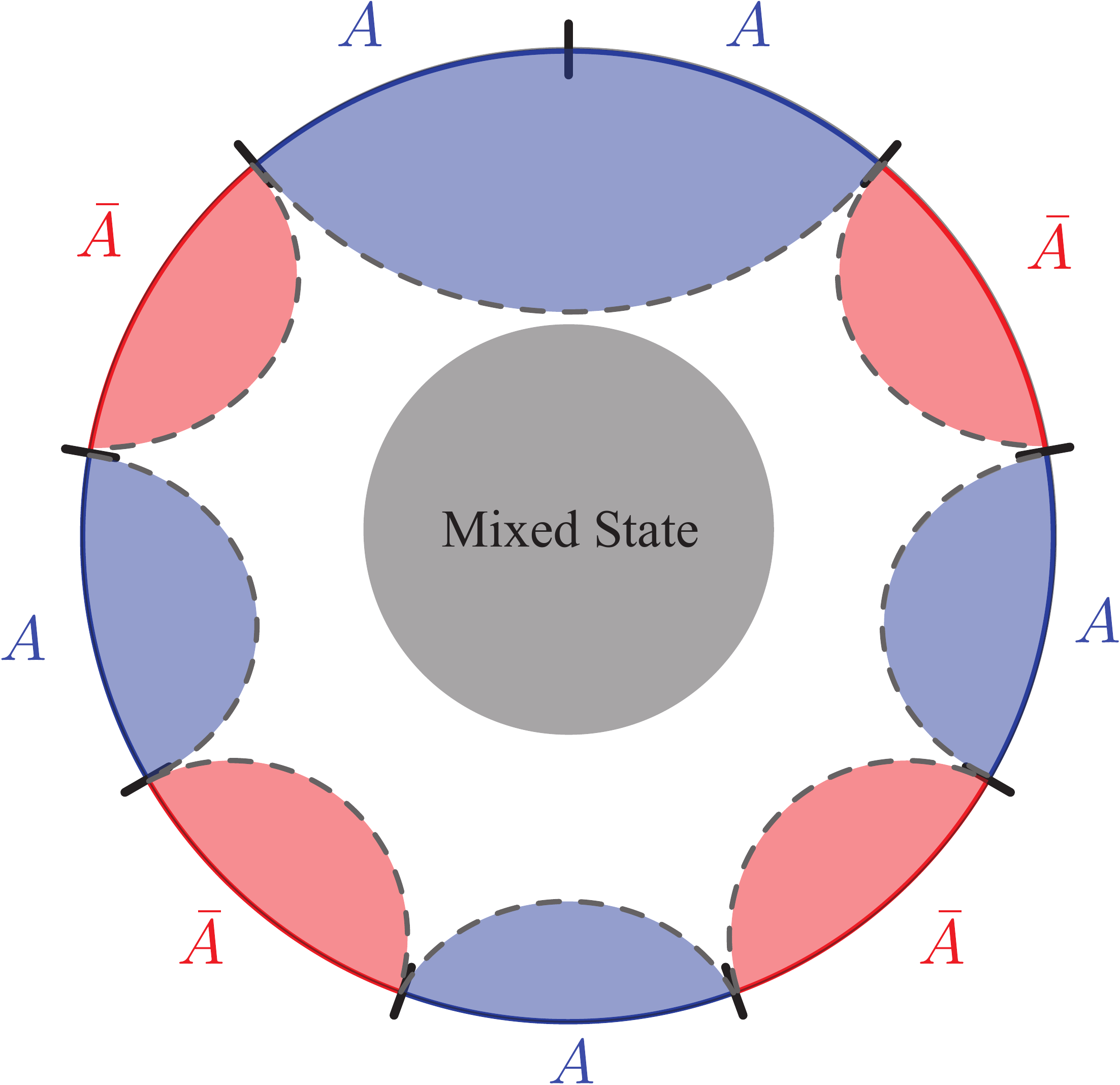}}
    \hfill
    \subfloat[]{\includegraphics[width=\columnwidth]{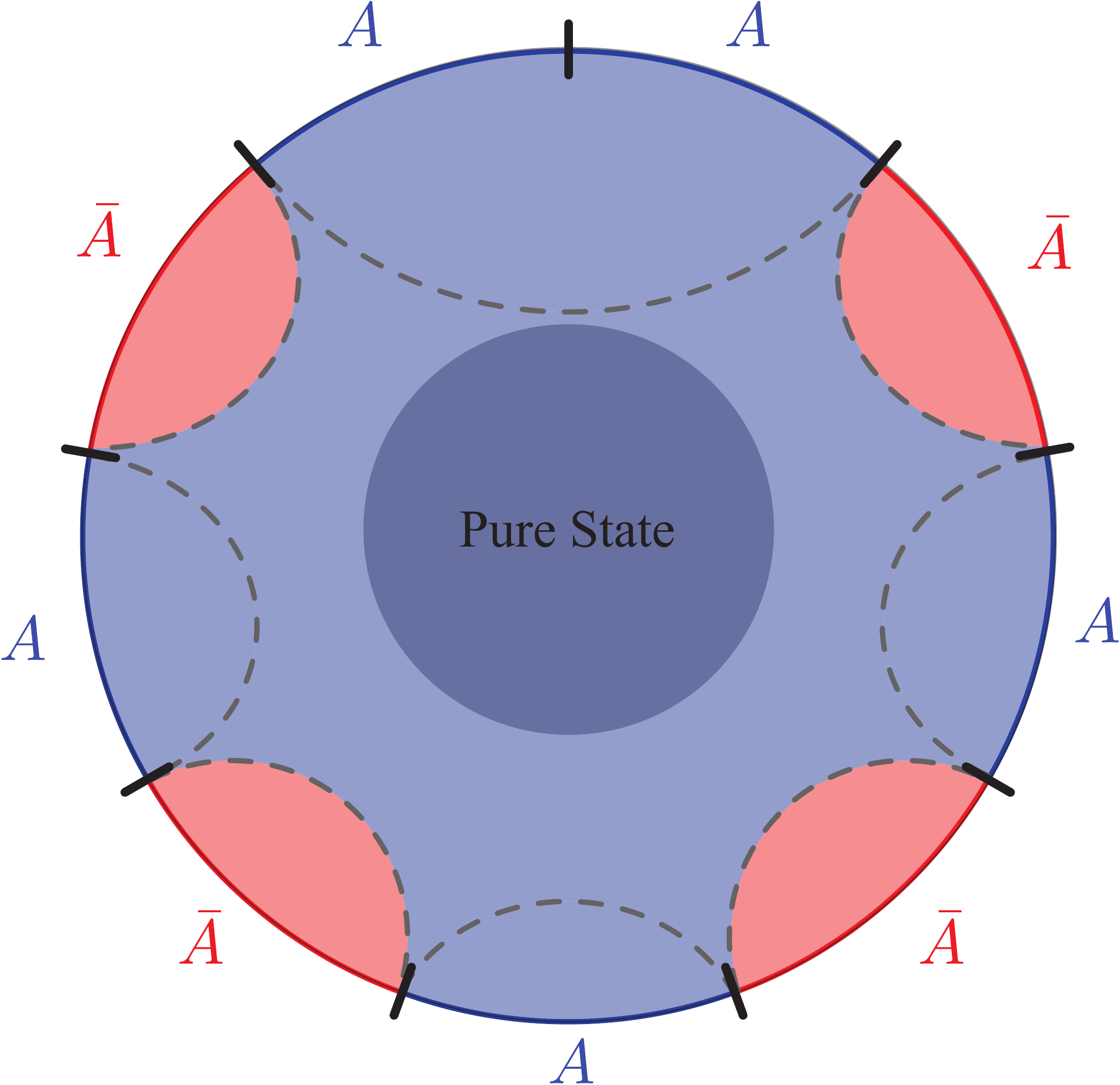}}
    \caption{In blue is the {\it entanglement} wedge of $A$ (blue boundary region). In red is the entanglement wedge of $\bar{A}$ (red boundary region). $A$ is defined by dividing the boundary into $2n+1$ equal pieces and taking the union of every-other connected component. The quantum extremal surface of each connected component is depicted as thin dashed gray lines. The dustball is the shaded circle in the middle.
    (a) Neither $A$'s nor $\bar{A}$'s entanglement wedge includes the dustball when it is in the maximally-mixed state in our code subspace. Therefore neither's reconstruction wedge includes the dustball. (b) $A$'s entanglement wedge includes the dustball when it is in a pure state. Therefore the entanglement wedge is much larger than the reconstruction wedge, in this state.}
    \label{fig:setup4and5}
\end{figure*}

The area of the extremal surface nearer to $A$ is greater than that nearer to its complement $\bar{A}$.
At leading order, this difference satisfies 
\begin{align}\label{eqn:deltaA}
    \frac{\Delta \mathcal{A}}{4 G_N} = \frac{\gamma \lads}{G_N}~,
\end{align}
where $\gamma$ is an $\mathcal{O}(1)$ constant independent of $n$.
The entropy of the dustball can easily be much greater than this: 
while the difference in areas is $n$-independent, the dustball entropy grows linearly with the radius $R$.
Hence by increasing $R$, we can make the entropy of the dustball dominate over $\Delta \mathcal{A}/4G_N$. 
Moreover, by increasing $n$ we can ensure the dustball remains safely outside the QES of any connected component of $A$ and $\bar{A}$.
Together, these properties ensure that the QES of both $A$ and $\bar{A}$ lie outside of the dustball when it is in the maximally-mixed state.
See Figure \ref{fig:setup4and5}(a).

On the other hand, when the spins of the dustball are in a pure state, $A$'s entanglement wedge will include the dustball.
See Figure \ref{fig:setup4and5}(b).
Therefore, such pure states are examples in which the reconstruction wedge of $A$ is much smaller than the entanglement wedge.
Indeed, by making the dustball arbitrarily large, one can make the reconstruction wedge smaller than the entanglement wedge by an arbitrarily large amount!

It is worth noting how the size and number of connected components of $A$ scale with $G_N$ in this example. As noted above, we need $n$ to scale proportional to $R$ in order to guarantee that the quantum extremal surfaces lie outside of the dustball, and this means that $n$ is proportional to the maximal entropy of the dustball. To find large separations between the EW and RW of $A$ we need this maximal entropy to be $\mathcal{O}(G_N^{-1})$, which means $n = \mathcal{O}(G_N^{-1})$, and hence the size of the intervals is $\mathcal{O}(G_N)$.
Perhaps it is an important feature of AdS/CFT that regions composed of such tiny pieces in general have trouble reconstructing their entanglement wedges.

\section{Discussion}
In this note, we have constructed a series of simple code subspaces in AdS/CFT where the reconstruction wedge differs macroscopically from the entanglement wedge in certain states, and thus there is a large portion of these entanglement wedges that cannot be reconstructed.
Our example signals the need for subspace-dependent reconstruction even in states where there are no horizons in the bulk. We hope that these examples prove to be more tractable than the black hole case. Ideally by studying cases where the reconstruction wedge is significantly inside the entanglement wedge, we can clarify the role of state-dependence in reconstructing the black hole interior.

We note that $2{+}1$d was not crucial for our analysis. 
For example, we expect that in even dimensions, for $A$ and $\bar{A}$ defined as unions of many strip-like regions (similar to what we have described, with each interval now extended infinitely in the new spatial dimensions), the leading-order area divergences in the CFT entropy will cancel between the near-$A$ and near-$\bar{A}$ extremal surfaces.
From the remaining log term, one still obtains a formula like equation \eqref{eqn:deltaA} in which $\Delta \mathcal{A}/4G_N$ is constant in the number of boundary strips and therefore can find dustball states for which the reconstruction wedge is arbitrarily small relative to the entanglement wedge.

\acknowledgements
It is a pleasure to thank Ning Bao, Raphael Bousso, Daniel Harlow, Patrick Hayden, Geoff Penington, Pratik Rath, and Arvin Shahbazi-Moghaddam for useful discussions.
X was formerly known as Google[x] and is part of the Alphabet family of companies, which includes Google, Verily, Waymo, and others (www.x.company).
This work was supported in part by the Berkeley Center for
Theoretical Physics, by the National Science Foundation (award number PHY-1521446), and by the U.S. Department of Energy under contract DE-AC02-05CH11231 and award
DE-SC0019380. The work of AL is supported by the Department of Defense (DoD) through the National Defense Science \& Engineering Graduate Fellowship (NDSEG) Program.

\bibliographystyle{utcaps}
\bibliography{all}

\end{document}